\documentclass{iaus}
\usepackage{graphicx}

\def\ms{M$_{\odot}$}
\def\zs{Z$_{\odot}$}

\title[Evolution of the Galaxy] %% give here short title %%
{On the chemical evolution of the Milky Way}

\author{N. Prantzos}   %% give here short author list %%
\affiliation{IAP, 98bis Bd Arago, 75013 Paris; email: {\tt prantzos@iap.fr}}

\pubyear{2008}
\volume{xxx}  %% insert here IAU Symposium No.
\pagerange{119--126}
\jname{Title of your IAU Symposium}
\editors{A.C. Editor, B.D. Editor \& C.E. Editor, eds.}
\begin{document}

\maketitle

\begin{abstract}
I discuss three different topics concerning the chemical evolution of the Milky Way (MW).
1) The metallicity distribution of the MW halo; it is shown that this distribution can be
analytically derived in the framework of the hierarchical merging scenario for galaxy formation,
assuming that the component sub-haloes had chemical properties similar to those
of the progenitors of satellite galaxies of the MW. 2) The age-metallicity relationship (AMR) in the
solar neighborhood; I argue for caution in deriving from data with important 
uncertainties (such as the age uncertainties in the Geneva-Kopenhaguen survey) a relationship between average metallicity and age: derived relationships are shown to be systematically flatter than
the true ones and should not be directly compared to models. 3) The radial mixing of stars in the 
disk, which may have important effects on  various observables (scatter in AMR, extension of the
tails of the metallicity distribution, flatenning of disk abundance profiles). Recent SPH +
N-body simulations find considerable radial mixing, but only comparison to observations
will ultimately determine the extent of that mixing.

\end{abstract}

%\section{Introduction} 
% if your document starts with a section,

\section{The MW halo in cosmological context}

 The regular shape of the metallicity distribution of the Milky Way (MW) halo can readily be explained by the simple model of galactic chemical evolution (GCE) with outflow, as suggested by Hartwick (1976). However, that explanation lies within the framework of the monolithic collapse scenario for the formation of the MW (Eggen, Lynden-Bell and Sandage, 1962).
Several attempts to account for the metallicity distribution of the MW halo  in the
modern framework (hierarchical merging of smaller components, hereafter sub-haloes) were undertaken in recent years,  (Bekki and Chiba 2001;  Scanapieco and Broadhurst 2001,   Font et al. 2006,  Tumlinson 2006,  Salvadori, Schneider and Ferrara 2007).  Independently of their success or failure in reproducing the observations, those recent models provide little or no physical insight into the physical processes that shaped the metallicity distribution of the MW halo. Indeed, if the halo was built from a  large number of sucessive mergers of sub-haloes, why is its metallicity distribution so well described by the simple model with outflow (which refers to a single system)? And what determines the peak of the metallicity distribution at [Fe/H]=$\sim$--1.6, which is (successfully) interpreted in the simple model by a single parameter (the outflow rate) ? 

In this work we present an attempt to built the halo metallicity distribution analytically, in the framework of the hierarchical merging paradigm. Preliminary results have already been presented in Prantzos (2007b) and a detailed account is given in Prantzos (2008). It is assumed that the building blocks were galaxies with properties similar to (but not identical with) those of the Local group dwarf galaxies that we observe today. In that way, the physics of the whole process becomes (hopefully) more clear.

\subsection{The halo metallicity distribution  and the simple model}

The halo metallicity distribution is nicely described by the simple model of GCE, in which the metallicity $Z$ is given as a function of the gas fraction $\mu$ as $ Z \ = \ p \ ln(1/\mu) \ + Z_0 $,
where $Z_0$ is the initial metallicity of the system  and $p$ is the {\it yield} 
(metallicities and yield are expressed in units of the solar metallicity
\zs).  If the system evolves at a
constant mass (closed box), the yield is called the {\it true yield},
otherwise (i.e. in case of mass loss or gain) it is called the {\it
effective yield}. The {\it differential metallicity
distribution} (DMD)  is:
\begin{equation} 
{{d(n/n_1)}\over{d(logZ)}} \ = \
 {{{\rm ln10}}\over{1-{\rm exp}\left(-{{Z_1-Z_0}\over{p}}\right)}}  \
 {{Z-Z_0}\over{p}} 
\ {\rm exp}\left(-{{Z-Z_0}\over{p}}\right)
\end{equation}
where $Z_1$ is the final metallicity of the system and $n_1$ the total
number of stars (having metallicities $\le Z_1$). This function has a
maximum for $Z-Z_0=p$, allowing one to evaluate easily the effective yield $p$ if the DMD is determined observationally.

The DMD of field halo stars peaks at [Fe/H]$\sim$--1.6, suggesting an effective halo yield 
 $p_{Halo} \sim$1/40 \zs \ for Fe.
This has to be compared to the true yield obtained in the solar neighborhood, which is $\sim$ \zs \
(from the peak of the local MD), but $\sim$2/3 of that originate in SNIa, so that
the true Fe yield of SNII during the local disk evolution
was $p_{Disk}\sim$0.32 \zs;
consequently, the {\it effective Fe yield} of SNII during the halo phase (where they
dominated Fe production) was $p_{Halo}\sim$0.08 $p_{Disk}$. 

 Hartwick (1976) suggested that {\it
outflow} at a rate $F \ = \ k \ \Psi$ (where $\Psi$ is the Star Formation Rate or SFR) occured
during the halo formation. In the framework of the simple model of GCE
such outflow reduces the true yield $p_{True}$ to its effective value $p_{Halo}=p_{True}(1-R)/(1+k-R)$
(e.g. Pagel 1997) where $R$ is the Returned Mass Fraction  ($R\sim$0.32-0.36 for most of the modern IMFs, e.g. Kroupa 2002).
The halo data suggest then that the ratio of the outflow to the star formation rate was 
\begin{equation}
k \ = \ (1-R) \ \left({\frac{p_{True}}{p_{Halo}}} -1 \right)
\end{equation}
and, by replacing $p_{True}$ with $p_{Disk}$ one finds that $k\sim$7-8. 

\begin{figure}
\centering
\includegraphics[width=0.49\textwidth]{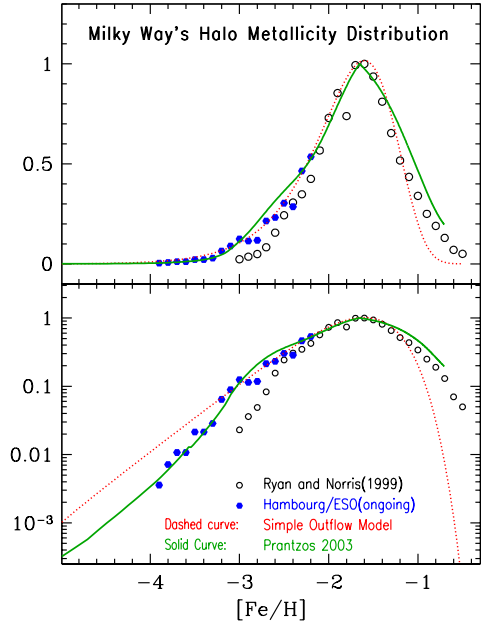}
\caption{Differential metallicity distribution of field halo stars in
 linear ({\it top}) and logarithmic ({\it bottom}) scales. Data are from
 Ryan and Norris (1991, {\it open symbols}), and the ongoing Hamburg/ESO 
 project  ( {\it filled
 symbols}). The two data sets are normalised at [Fe/H]-2.2; above that
 value, the Hamburg/ESO data are incomplete, while below that
 metallicity the Ryan and Norris (1991) data set is incomplete. 
The {\it dotted curve} is a simple model with instantaneous recycling (IRA) and  
outflow rate equal to 8 times the star formation rate. 
The solid curve is obtained as in Prantzos (2003), from a model without
IRA, an early phase of rapid infall and a constant outflow rate equal to
 7 times the SFR. All curves and data are normalised to max=1.}
%The histogram is the result of an early model (Bekki and Chiba 2001) 
%calculated in the modern framework of hierarchical merging (see text).    }
\label{fig:1}       % Give a unique label
\end{figure}

Although the siple model of GCE with outflow fits extremely well the
bulk of the halo DMD, the situation is less satisfactory for the low
metallicity tail of that DMD. 
Prantzos (2003), noting that the
situation is reminiscent of the local G-dwarf problem,  suggested that a similar solution should
 apply, namely an early phase of rapid infall (in a time scale
of less than 0.1 Gyr) forming the Milky Way's halo, as 
illustrated in Fig. 1. 

\subsection{The halo DMD and hierarchical merging}

In this work we assume that the MW halo was formed by the merging of
smaller units ("sub-haloes"), as implied by the hierarchical merging scenario 
for galaxy formation. In order to calculate semi-analytically
the resulting DMD, it is further  assumed that each of the sub-haloes
had a DMD described well by the simple model, i.e. by Eq. (1.1). It
remains then to evaluate the effective yield $p(M)$ of
each sub-halo as a  function of its mass, as well as 
the mass function of the sub-haloes $dN/dM$. 

It is assumed here that each one of the merging sub-haloes has a DMD
described by the simple model with an appropriate effective yield. This
assumption is based on recent observations of the dwarf spheroidal
(dSph) satellites of the Milky Way, as will be discussed below. It is
true that the dSphs that {\it we see today} cannot be the components of
the MW halo, because of  their observed abundance patterns (e.g. Shetrone, C\^ot\'e and Sargent 2001;
Venn et al. 2004): their $\alpha$/Fe ratios are typically smaller than
the   [a/Fe]$\sim$0.4$\sim$const. ratio of halo stars. 
This implies that they evolved on longer timescales than the Galactic
halo, allowing SNIa to enrich their ISM with Fe-peak nuclei and thus 
to lower the $\alpha$/Fe ratio by a factor of $\sim$2-3 (as evidenced from the 
[O/Fe]$\sim$0 ratio in their highest metallicity stars). 
However, the shape of the DMD of the
simple model does not depend on the star formation history or the
evolutionary timescale, only on gas flows into and out of the system, as well as on the
initial metallicity (e.g. Prantzos 2007a).

\begin{figure}
%\centering
\includegraphics[width=0.46\textwidth]{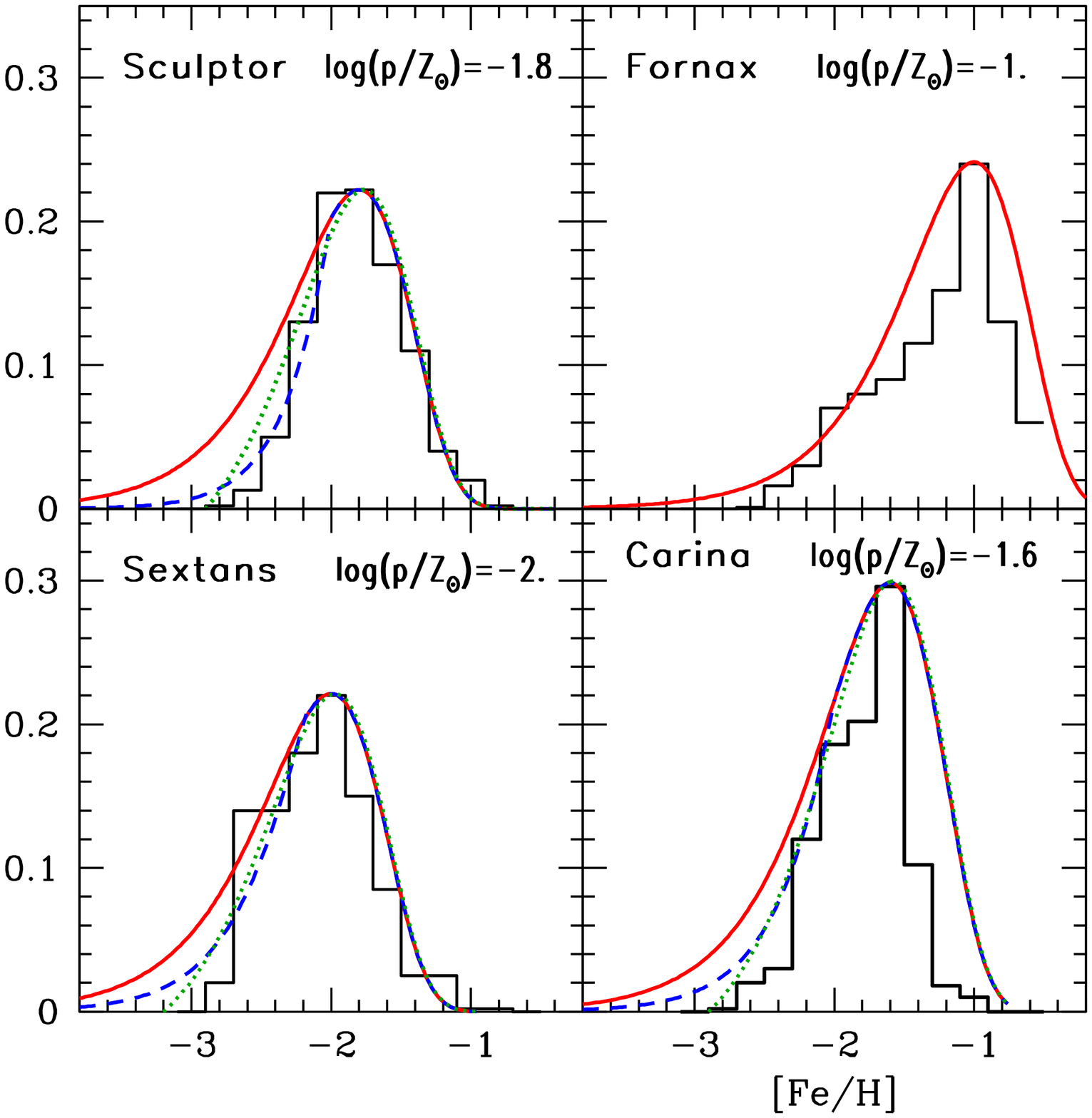}
\qquad
\includegraphics[width=0.46\textwidth]{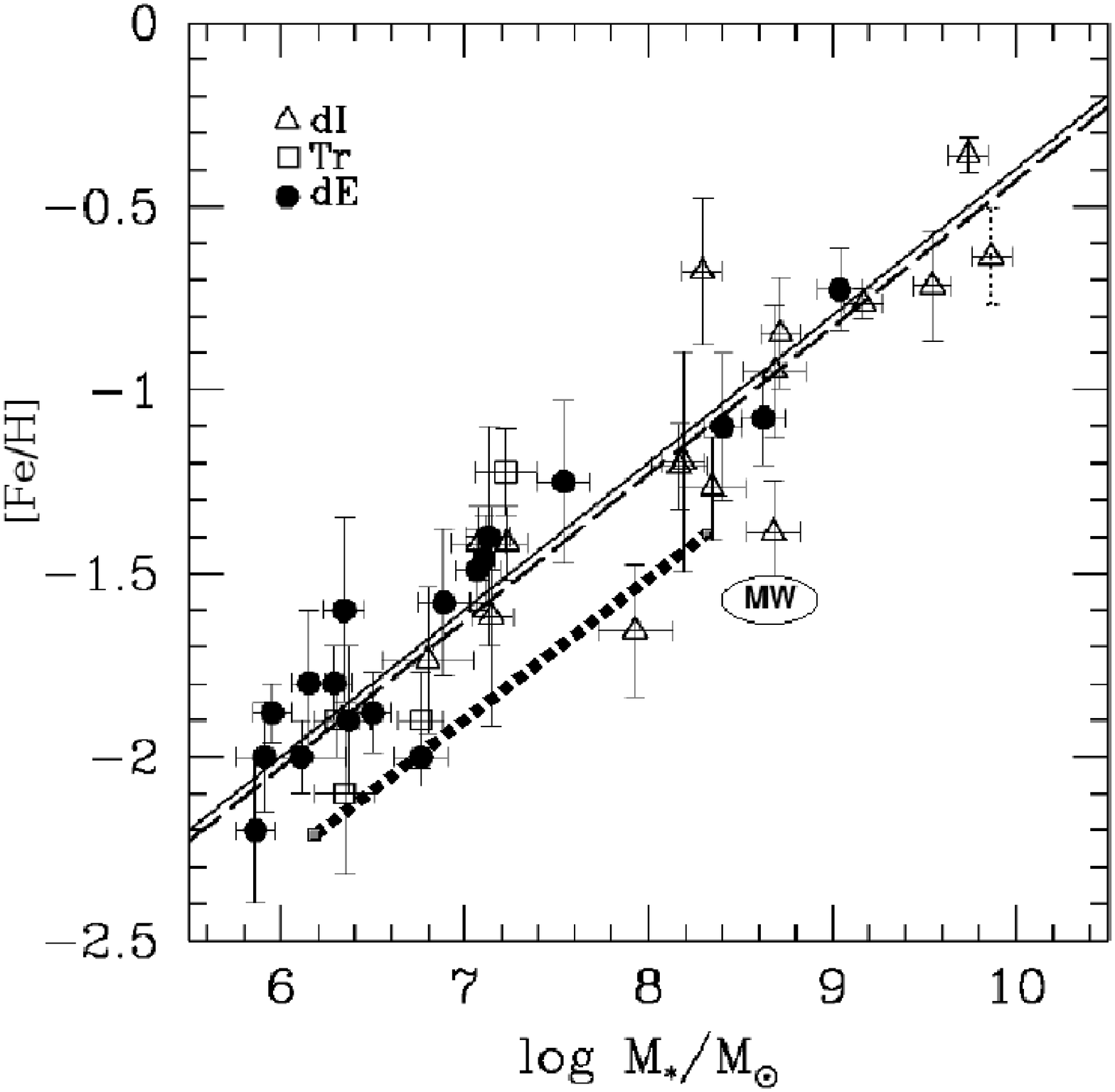}
\caption{{\bf left:} Metallicity distributions  of dwarf satellites of the Milky Way. 
Data are in {\it histograms} (from Helmi et al. 2006). {\it Solid curves} 
indicate the results of simple GCE  models with outflow proportional to 
the star formation rate;  the corresponding effective yields (in \zs) 
appear on top right of each panel. {\it Dashed curves} are fits obtained
 with an early infall phase, while {\it dotted curves} are models with
 an initial metallicity log$(Z_0)\sim$--3; both modifications 
to the simple model
 (i.e. infall and initial metallicity) improve the fits to the data.
{\bf Right:} Stellar metallicity vs stellar mass for nearby galaxies; data
 and model ({\it upper curves}) are from Dekel and Woo (2003), with {\it
 dI} standing for dwarf irregulars and  {\it dE} for dwarf ellipticals. The {\it thick 
dotted} line represents the effective yield of the sub-haloes that formed the MW halo according to this work (i.e. with no contribution from SNIa, see Sec. 3.2). The MW halo, with average metallicity [Fe/H]=--1.6 and 
estimated mass $\sim$4x10$^8$ \ms \ falls below both curves.  }
\label{fig:2}       % Give a unique label
\end{figure}

The DMDs of four neraby dSphs (Helmi et al. 2006) are displayed as histograms in Fig. 2,
where they are compared to the simple model with appropriate effective
yields ({\it solid curves}). The effective yield in each case was simply
assumed to equal the peak metallicity (Eq. 1.1). It can be seen that the overall shape 
of the DMDs is quite well fitted by the simple models. This is important,
since i) it strongly suggests that {\it all} DMDs of small galaxian
systems can be described by the simple model  and ii) it allows to determine {\it
effective yields} by simply taking the peak metallicity of each DMD (see below).

Before proceeding to the determination of effective yields, we note that the fit of the simple model to the data of dSphs fails in the low metallicity tails. Helmi et al. (2006) attribute this  to a pre-enrichment of the gas out of which the dSphs were formed. However, early infall is another, equally plausible, possibility, as argued in Prantzos (2008). The recent simulations of Salvadori et al. (2008) find both pre-enrichment and early infall for local dSphs.

If the DMDs of all the components of the Galactic halo are described by
the simple model, then their shape is essentially  described by the
corresponding effective yield $p$ (and, to a lesser degree, by the
corresponding initial metallicity $Z_0$). Observations suggest that the
effective yield is a monotonically increasing function of the galaxy's
stellar mass $M_*$ (Fig. 3).  In the case of the
progenitor systems of the MW halo, however, the effective yield must have been lower, since SNIa had not time to contribute (as evidenced by the high 
$\alpha$/Fe$\sim$0.4 ratios of halo stars), by a factor of about 2-3. We assume
then that the effective yield of the MW halo components (accreted satellites
or sub-haloes) is given (in \zs) by
\begin{equation}
p(M_*) \ = \ 0.005 \ \left({\frac{M_*}{10^6 M_{\odot}}}\right)^{0.4}
\end{equation}
i.e. the thick dotted curve in Fig. 3.

Obviously, the stellar mass $M_*$ of each of the sub-haloes should be $M_*< M_H$
where $M_H$ is the stellar mass of the MW halo ( $M_H$=
4$\pm$0.8 10$^8$ \ms, e.g. Bell et al. 2007).

Hierarchical galaxy formation scenarios predict  the mass function of
the dark matter sub-haloes which compose a dark matter halo at a given
redshift. Several recent simualtions find
$ dN/dM_D \ \propto \ M_D^{-2}  $ (Diemand et al. 2007, Salvadori et al. 2007,   Giocoli et al. 2008).
In our case, we are interested in the mass function of the
{\it stellar sub-haloes}, and not of the dark ones. Considering the effects of outflows on
the baryonic mass function, Prantzos (20008) finds that
\begin{equation}
\frac{dN}{dM_*} \ \propto \ M_*^{-1.2}
\end{equation}
i.e. {\it the distribution function of the stellar sub-haloes is flatter than
the distribution function of the dark matter sub-haloes}. The normalisation of the stellar sub-halo
distribution function is made through
\begin{equation}
\int_{M_1}^{M_2} \frac{dN}{dM_*} \ M_* \ dM_* \ = \ M_H
\end{equation}

\begin{figure}

\includegraphics[width=0.46\textwidth]{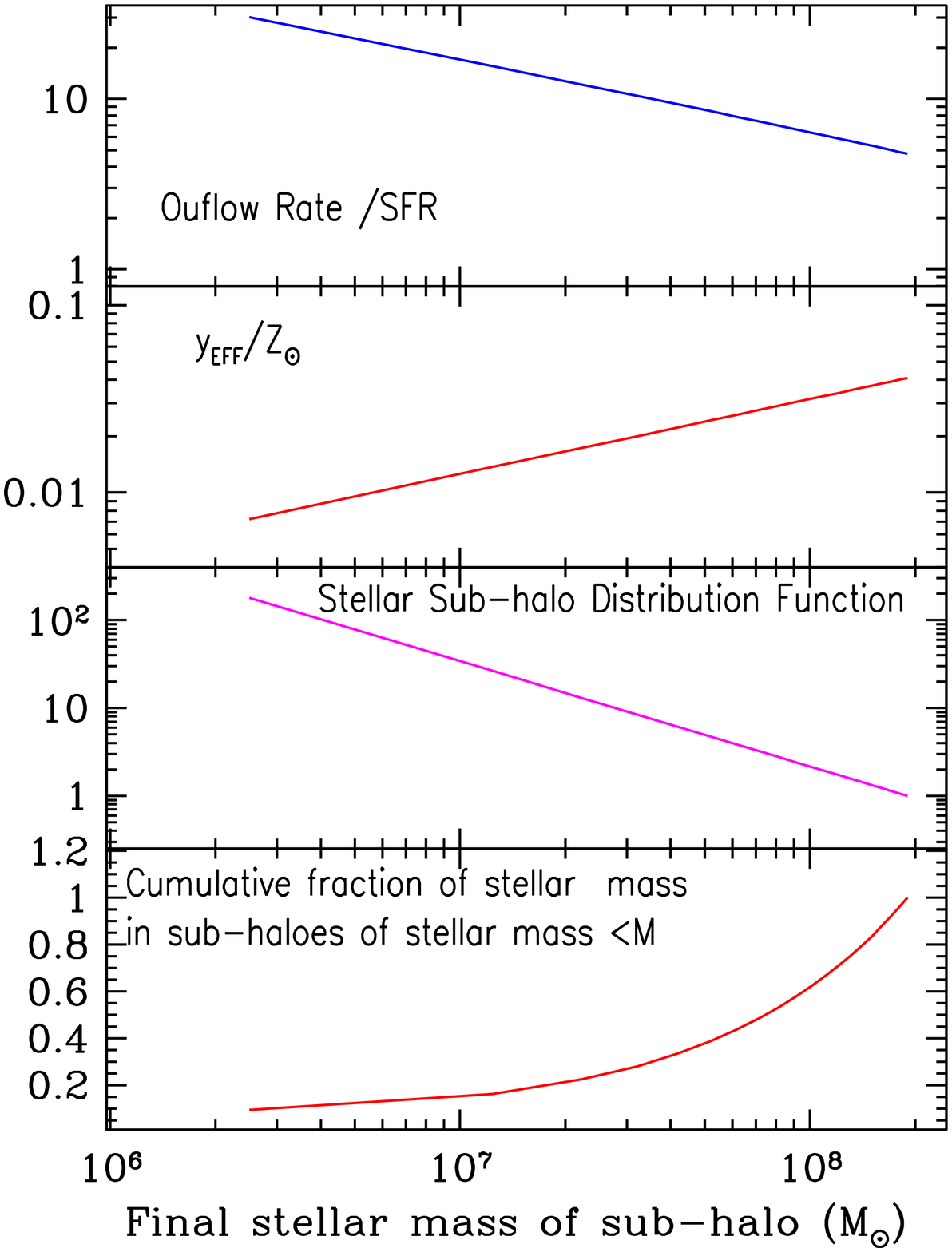}
\qquad
\includegraphics[width=0.46\textwidth]{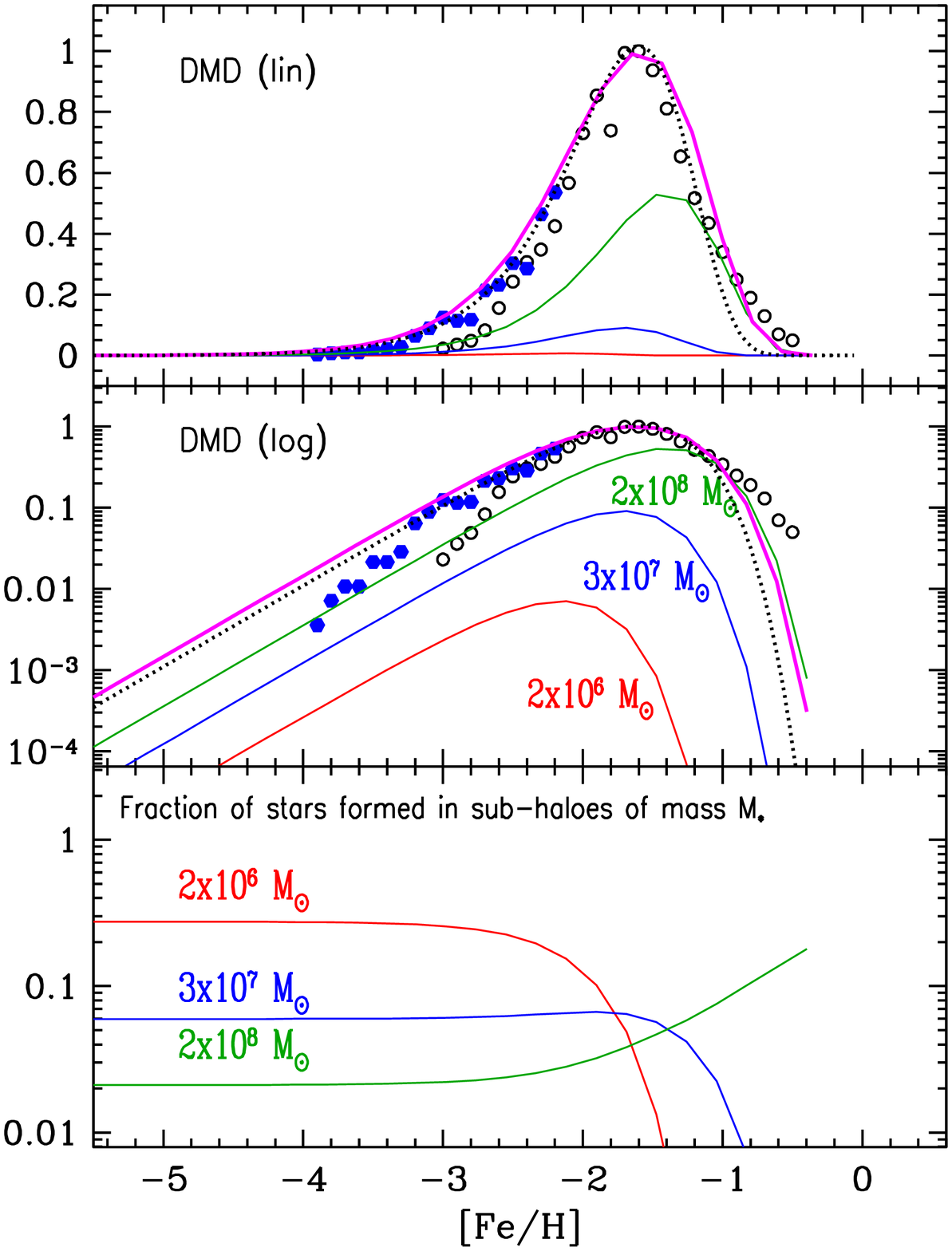}
\caption{{\bf Left:}Properties of the sub-haloes as a function of their stellar
 mass,  empirically derived as discussed
 in Sec. 3. From top to bottom: Outflow rate, in units of the
 corresponding star formation rate; Effective yield, in solar units;
 Distribution function; Cumulative fraction of stellar mass contributed
 by the sub-haloes. The total mass of the MW halo is 4 10$^8$ \ms.  {\bf
 Right}:{\it Top} and {\it middle} panels: Differential metallicity distribution (in lin and log scales, respectively) of the MW halo, assumed to be composed of a population of smaller units (sub-haloes). The individual DMDs of a few sub-haloes, from 10$^6$ \ms \ to 4 10$^7$ \ms, are indicated in the middle panel, as well as the sum over all haloes ({\it solid upper curves} in both panels, compared to observations). {\it Dotted curves} in top and middle panels indicate the results of the simple model with outflow (same as in Fig. 1). Because of their large number, small sub-haloes with low effective yields contribute the largest fraction of the lowest metallicity stars, while large haloes contribute most of the high metallicity stars ({\it bottom panel}).  }
\label{fig:4}
\end{figure}

The lower mass limit $M_1$ is adopted here to be $M_1$=1-2 10$^6$ \ms, in
agreement with the lower mass bound of observed dSphs in the Local group.
 Such galaxies have internal velocities $V>$10
km/s. Dekel and Woo (2003) argue that the gas in haloes with
$V<$10 km/s cannot cool to form stars at any early epoch and that dwarf
ellipticals form in haloes with 10$<V<$30 km/s, the upper limit corresponding to 
a stellar mass $M_*\sim$2 10$^8$ \ms. This is the 
upper mass limit $M_2$ that we adopt here. 
The main properties of the sub-halo set constructed in this section
appear in Fig. 3 (left) as a function of the stellar sub-halo mass $M_*$.
The resulting total DMD is obtained as a sum over all sub-haloes:
\begin{equation}
\frac{d(n/n_1)}{d(logZ)} \ = \ \int_{M_1}^{M_2} \ \frac{d[n(M_*)/n_1(M_*)]}{d(logZ)} \ \frac{dN}{dM_*} \ M_* \ dM_*
\end{equation}
The result appears in Fig. 4 (right, with top panel in linear and middle panel in logarithmic scales, respectively). It can be seen that it fits the observed DMDs at least as well as the simple model \`a la Hartwick. 
In summary, under the assumptions made here, the bulk of the
DMD of the MW halo results naturally as the sum of the DMDs of the
component sub-haloes. It should be noted that all the ingredients of the analytical model are taken from observations of
local satellite galaxies, except for the adopted mass function of the sub-haloes (which results from analytical theory of structure formation plus a small modification to 
account for the role of outflows). Obviously, by assuming different values 
for the slope of the dark matter haloes  and for the mass limits $M_1$ and $M_2$ 
one may modify the peak of the resulting composite DMD, thus allowing for differences 
between the halo DMDs of different galaxies.

\section{The local age-metallicity (AMR) relationship}

In  the framework of the simple model of galactic chemical evolution there is a unique relationship between the abundance of a given element/isotope and time. 
The solar neighborhood is at present the only galaxian system where the age-metallicity relation (AMR) can be measured. Starting with the pioneering works of Mayor (1974) and Twarog (1980), the local AMR has been extensively studied over the years with larger and better defined samples (see e.g. Feltzing et al. 2001 for references); the aim of those studies is to find the shape of the AMR and 
whether it displays intrinsic scatter, i.e. exceeding observational errors. The existence of such a scatter was strongly supported by the  work of Edvardsson et al. (1993), who used accurate (spectroscopically determined) metallicities for a sample of 163 stars. Needless to say that simple models do not, by construction, allow for scatter, unless complementary assumptions are made.

The most comprenhesive recent study of the stars in the solar neighborhood is the Geneva-Kopenhaguen Survey (GKS), presented in Nordstrom et al. (2004) and updated in Holmberg et al. (2007). It concerns a magnitude limited sample of more than 13000 F and G dwarfs with accurately determined distances and kinematics; metallicities are photometrically determined and have estimated errors $\sim$0.1 dex, while stellar ages suffer from considerable uncertainties, often excceding 50\%.
The AMR determined by using the large sample, or various sub-samples (to account for different types of biases) displays 
an upturn for stars of young ages and solar metallicities (which dominate largely the sample), but is essentially flat otherwise (see upper left panel of Fig. 4). Similar results (i.e. an essentially flat AMR, when {\it average metallicity} is calculated as a function of age) are obtained in the independent study of Subiran et al. (2007) with a sample of $\sim$700 giant stars (data points with errors bars in the same figure).

\begin{figure}
\includegraphics[width=0.46\textwidth]{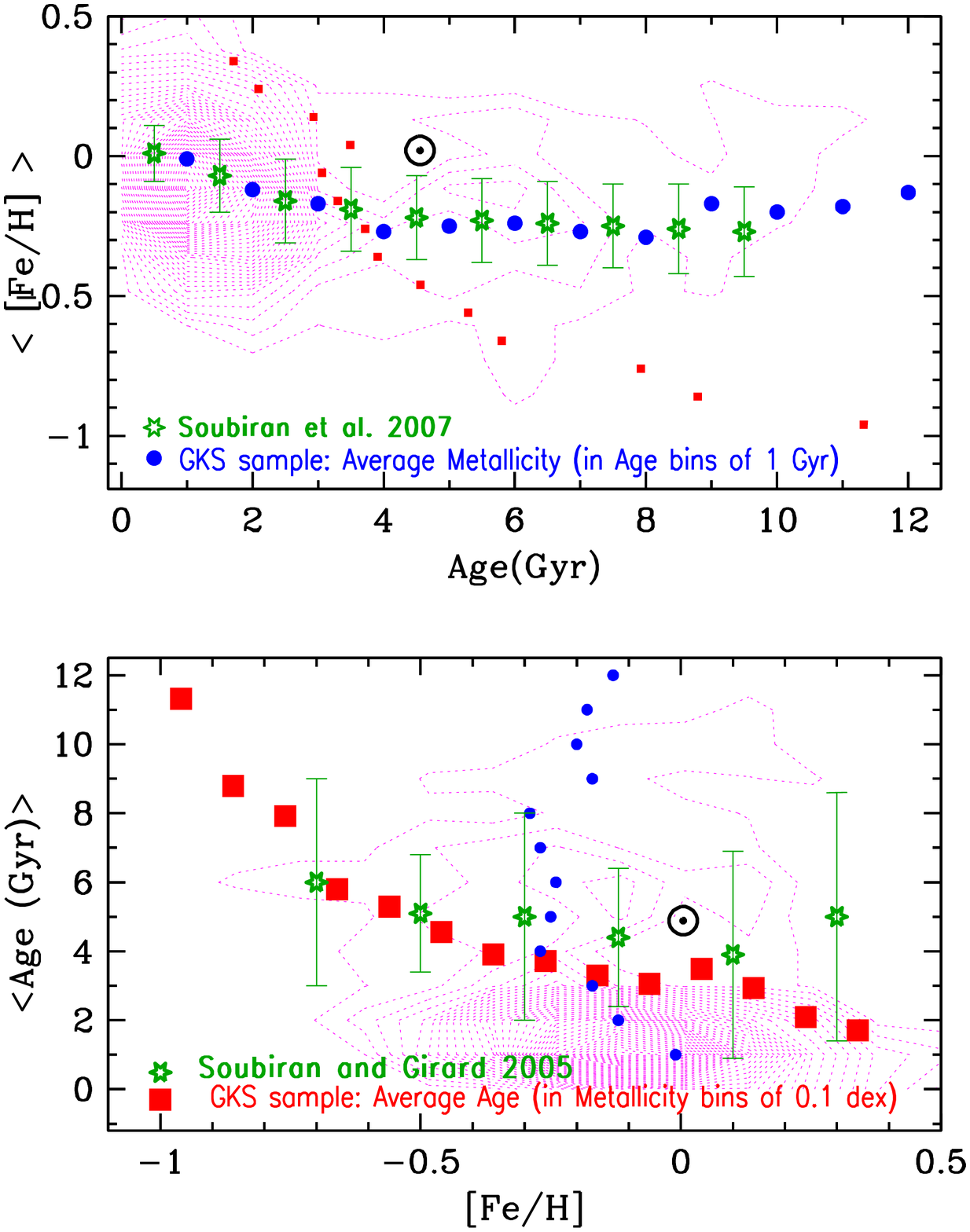}
\qquad
\includegraphics[width=0.46\textwidth]{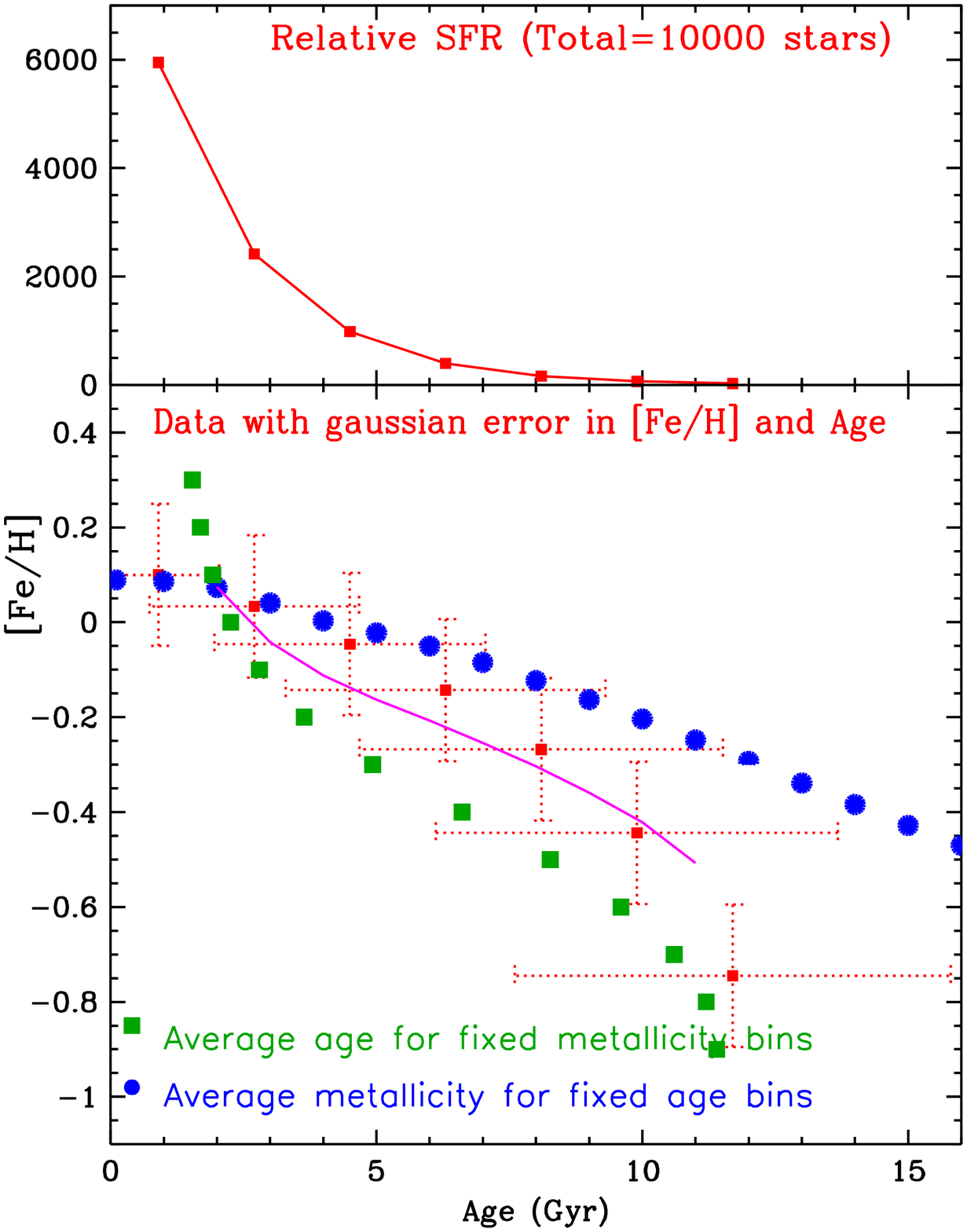}
\caption{{\bf Left:} Local Age-Metallicity relationship, in 1 Gyr age bins ({\it top}) 
and in 0.1 dex metallicity bins ({\it bottom}); {\it isocontours} correspond to the data 
of GKS (Nordstrom et al. 2004), {\it large symbols} show the corresponding relationships 
of age vs mean metallicity ({\it top}) and Metallicity vs mean age ({\it bottom}), 
{\it small symbols} the complementary relationship,  and data with error bars designate 
results of independant studies with different samples (from Subiran  et al. 2007 and 
Subiran and Girard 2005). Notice that in both panels the Sun appears to be slightly
metal-rich for its age or slightly older than stars of the same metallicity.
{\bf Right:} By using simulated data as input for the AMR (points with 1-$\sigma$ 
Gaussian error bars in lower panel, as indicated by {\it dotted lines}) and a star 
formation rate  exponentially increasing with time (top panel), one may derive the corresponding
age vs average metallicity ({\it circles}) and metallicity vs average age ({\it squares})
relations; the former is flatter than 
and the latter is steeper than the input relationship (as with the case of the real 
data, see left pannels). Their average ({\it solid curve}) is a better approximation
to the input data.}
\end{figure}

If the AMR is indeed as flat as suggested by those studies, this would have important implications for our understanding of the evolution of the solar neighborhood: taken at face value, it would imply that the local Galaxy evolved essentially at constant average metallicity for most of its lifetime ($\sim$0.5 \zs), with a small increase in the past few Gyr. None of the present-day models, which satisfy an important number of other observational constraints (e.g. Boissier and Prantzos 1999, Goswami and Prantzos 2000), predict such an 
evolution; in general, metallicity increases early on and saturates at late times.

Before getting to conclusions based on the AMR one should seriously consider the  various biases affecting it. That issue is properly emphasized in Nordst\" rom et al. (2004), Haywood (2006), Holmberg et al. (2007). Here we wish to draw attention to an aspect of the AMR which is rarely considered, despite the fact that the quantitative determination of a correlation between two properties of a sample of objects is one of the long standing problems in observational astronomy (and in many other fields of science).

Given the two properties (metallicity and age) of the sample stars, one may ask either what is the {\it average metallicity at  a given age } or {\it what is the average age at a given metallicity}. If there were a unique AMR (with no scatter, as predicted by the simple model) the answer would be obviously the same in both cases. However, in the presence of scatter in the data, {\it the answer  is not the same}. This can be seen in the bottom left panel of Fig. 4, where average age is calculated for fixed metallicity bins of 0.1 dex ({\it filled squares}): there is substantial variation of the average age with metallicity in this case, at variance with the average metallicity vs age relationship derived before. Notice that Subiran and Girard (2005), using a different sample, find a flatter $<$age$>$ vs metallicity relationship (data points with error bars in lower left panel of Fig. 4), but again orthogonal to the $<$metallicity$>$ vs age relationship.

In order to better understand that situation, we performed calculations with artificial input data and one of the results is displayed on the right panel of Fig. 4. We adopted a SFR increasing with time (upper right panel in Fig. 4), to simulate the large number of young stars in the GKS sample and we normalised the total number to 10000 stars (comparable to the size of the GKS sample). We adopted an average AMR steadily increasing with time
(small points in bottom right panel) with Gaussian scatter of 0.1 dex in metallicity and (age/Gyr) in age, as indicated by the {\it dotted error bars}. The resulting $<$age$>$ vs metallicity and $<$metallicity$>$ vs age 
relationships are also displayed in the same figure ({\it thick squares}  and {\it dots}, respectively).
They differ considerably from each other and from the input data, converging at young ages, where the number of stars is large. Overall, the figure shows considerable overall similarity with
real data analysis on the left. We checked that by reducing the error bars the two derived relationships converge everywhere to the input AMR, as they should, and we tested different SFR with the same input AMR (not displayed in Fig. 4). It turns out that in the case of a strong early SFR the two curves converge at old ages and diverge at young ages, contrary to the case displayed here. The situation is intermediate in the case of a constant SFR.

What do we learn from this analysis? That in the presence of scatter in the data, the derivation of the AMR becomes a delicate enterprise. Of course, if the scatter is intrinsic (due to some kind of inhomogeneous chemical evolution) the concept of $<$metallicity$>$ vs age is more physicaly meaningful than the one $<$age$>$ vs metallicity and this explains why in most cases only  the former relationship is calculated from the data. However, in case of important 
scatter in the data, such as the one affecting the ages of the GKS sample, the former relationship does not reflect the true AMR, as shown here with simulated data (which also reproduce convincingly  the real situation). In that case, the true AMR could, perhaps, be better approximated by taking some arithmetic mean between the two derived relationships (as illustrated by the {\it thin curve} in the bottom right panel of Fig. 4). We advance this with caution, since we have not yet thorougly checked this idea; we simply notice that it is not very different from the ordinary least squares bisector method, which is often used in astronomy, especially after  Isobe et al. (1990) presented a convincing study showing its advantages over other data  analysis methods.

\section{Radial mixing in the Milky way disk}
The idea that stars in a galactic disk may diffuse to large distances along the radial 
direction (i.e to distances larger than allowed by their epicyclic motions) was proposed 
by Wielen et al. (1996). They suggested that some of the peculiar chemical properties 
of the Sun may be explained by the assumption that it was born in the inner Galaxy 
(i.e. in a high metallicity region, in view of the galactic metallicity 
gradient) and subsequently migrated outwards. They treated the hypothetical radial 
migration phenomenologically, acknowledging that the basic mechanism for the gravitational perturbations of stellar orbits is not understood.

Sellwood and Binney (2002, herefter SB02) convincingly argued that stars can 
migrate over large radial distances,
due to resonant interaction with spiral density waves at corotation. Such a migration alters 
the specific angular momentum of individual stars, but affects very little the overall 
distribution of angular momentum and thus does not induce important radial heating of the disk.
Because high-metallity stars from the inner (more metallic and older) and the outer 
(less metallic and younger) disc are brought in the solar neighborhood, SB02 showed 
with a simple toy model that considerable scatter may result in the local age-metallicity
relation, not unlike  the one observed by Edvardsson et al. (1993). 

Another obvious implication of the radial migration model of SB02 concerns the flattening 
of the stellar metallicity gradient in the galactic disk. That issue was quantitatively 
explored in Lepine et al. (2003), who  considered, however, the corotation at a fixed radius 
(contrary to SB02). As a result, the gravitational interaction bassically removes stars 
from the local disk, "kicking" them inwards and outwards. The abundance profile 
(assumed to be initially exponential) is little affected in the inner Galaxy, but some
 flattening is obtained in the 8-10 kpc region. The authors claim that such a 
 flattening is indeed observed (using data of planetary nebulae by Maciel and Quireza 1996)
but modern surveys do not find it (see e.g. Cunha et al. this meeting).

\begin{figure}
\includegraphics[width=0.46\textwidth]{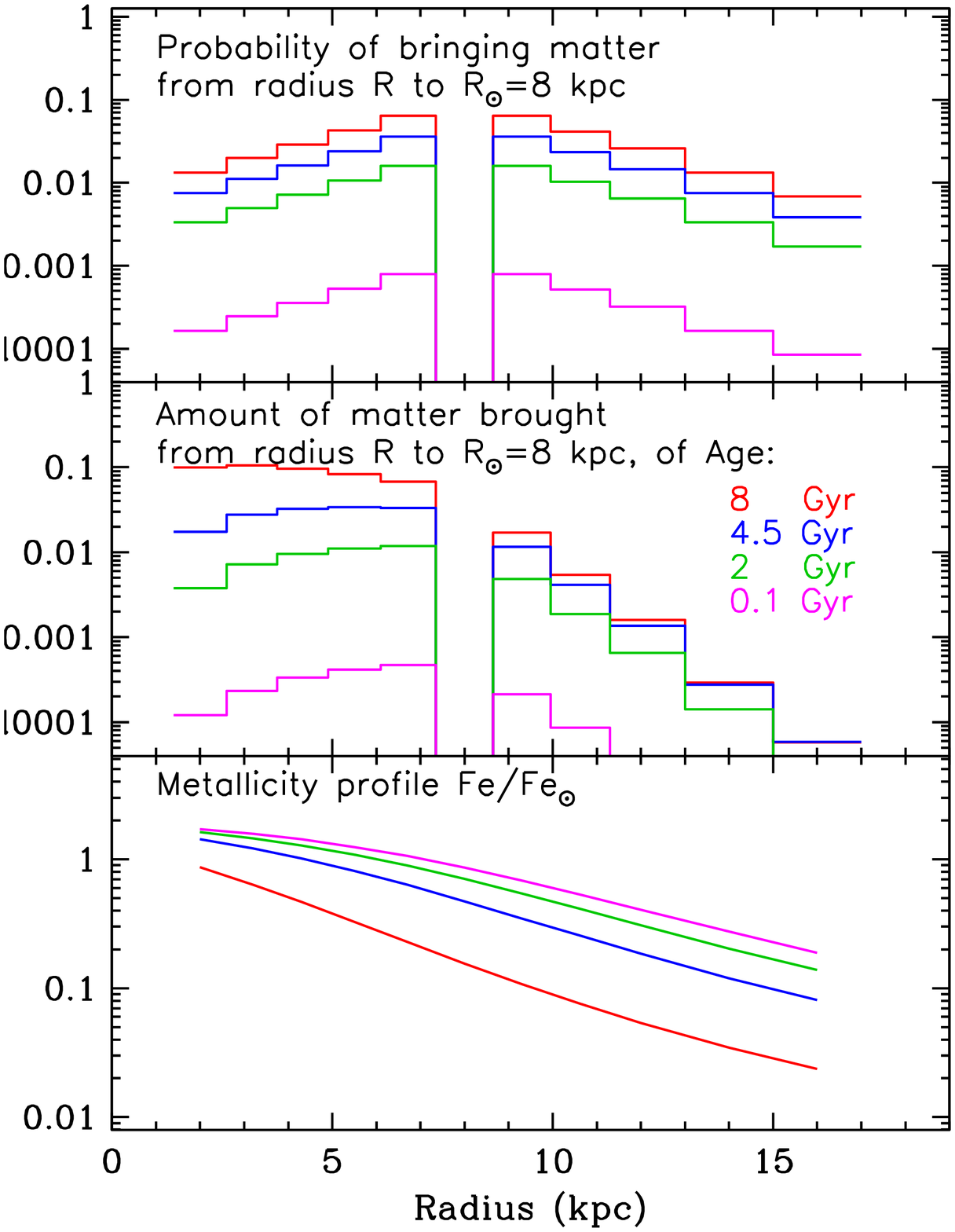}
\qquad
\includegraphics[width=0.46\textwidth]{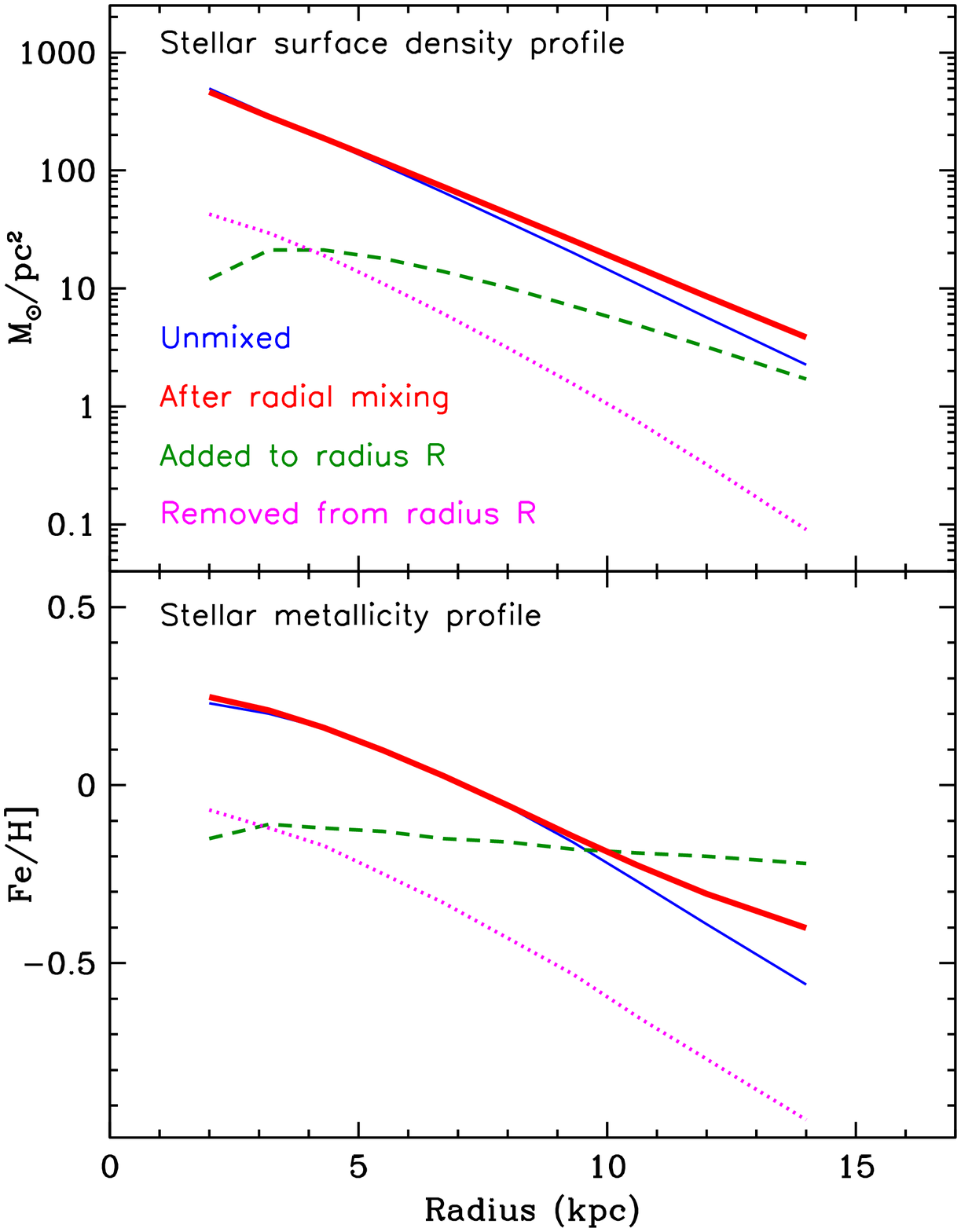}
\caption{{\bf Left:} A toy model of radial mixing, applied to a " realistic" disk (as obtained in Boissier and Prantzos 1999).
The various histogrammes/curves correspond to material of different ages (8, 4.5, 2 and 0.1 Gyr, respectively, from
top to bottom in the {\it top} and {\it middle} panels and from bottom to top in the {\it bottom} panel). {\bf Right:}
impact of radial mixing on stellar and metallicity profiles of the disk. {\it Solid curves} indicate the full profiles
before ({\it thin}) and after ({\it thick}) mixing: the external regions find their surface density and metallicity enhanced after mixing. {\it Dashed } and {\it dotted} curves in both panels indicate material added and removed, respectively from a given zone. }
\end{figure}

Finally, Haywood (2008), on the basis of kinematics and abundance observations of a 
large sample of local stars argues that most of the metal rich stars in the solar
neighborhood originate from the inner disk and most of the metal poor ones from the outer disk, 
and suggests that the local disk started its evolution with a considerably high metallicity
of [Fe/H]$\sim$-0.2. Such a large pre-enrichment of the thin disk, however, is difficult
to accept. Haywood (2008) attributes it to the thick disk, but despite the
relatively large contribution ($\sim$20\%) of that component to the local surface density, 
it cannot have pre-enriched to such a high degree the much more massive thin disk, assuming that the latter evolved as a closed box (as argued in Haywood 2006). And if infall is invoked for the thin disk
(despite the fact that  is superfluous  for the G-dwarf problem if pre-enrichment is assumed), then a large dilution of the initially assumed [Fe/H]$\sim$-0.2 would result. Independantly, however,
of his far-reaching conclusions, Haywood (2008) presents convincing arguments that the local 
stellar population shows evidence for radial migration.

We study some of the implications of radial migration with a toy model similar to the one of
SB02, but with boundary conditions based on a "realistic" simple model for the Milky 
Way disk, which reproduces most of the available observations (Boissier and Prantzos 1999). The key 
quantity is the probability that a star migrates   radially and we adopt here an
 function decreasing with distance
(remote stars have smaller probability of being found in the solar neighborhood that nearby ones)
and time (older stars have motre time and thus  higher probability of being found here 
than younger ones). Although the probabilities are intrinsically symmetric in the radial
direction for a given radius (Figure 5, top left),
the larger surface density of the inner disk results in larger 
numbers of stars transferred from the inner disk at 8 kpc than from the outer one 
(Fig. 5 middle left). 

We adopt (somewhat arbitrarily) a normalisation of the probability function of radial migration, as to not affect the stellar surface density by more than 10\% locally (at $R$=8 kpc). 
This is, perhaps, an underestimate
(as discussed below) but the toy model is presented here only for illustration purposes.
In Fig. 5 (top right) it is seen that with this normalisation,
regions at $R>$4 kpc receive more material (dashed curve) than it is removed from them (dotted curve). 
The exponential stellar profile of the original model
(thin curve) flattens somewhat outwards, with the stellar density almost being doubled
(thick curve)  at $R$=14 kpc.

The mean metallicity of all the stars received at a 
given region is always larger than the average metallicity of the stars ever removed 
from that region (dashed and dotted curves, 
respectively, in Fig. 5 bottom right)  
{\it but not from the original final average (over all stars) metallicity of that
same region, except for the zones at} $R>$10 kpc (thin solid curve in Fig. 5 right bottom). 
The average 
metallicity of these zones is affected by the migrated stars and it increases
by a factor of 2 (0.3 dex) at  $R$=14 kpc.

\begin{figure}
\includegraphics[width=0.46\textwidth]{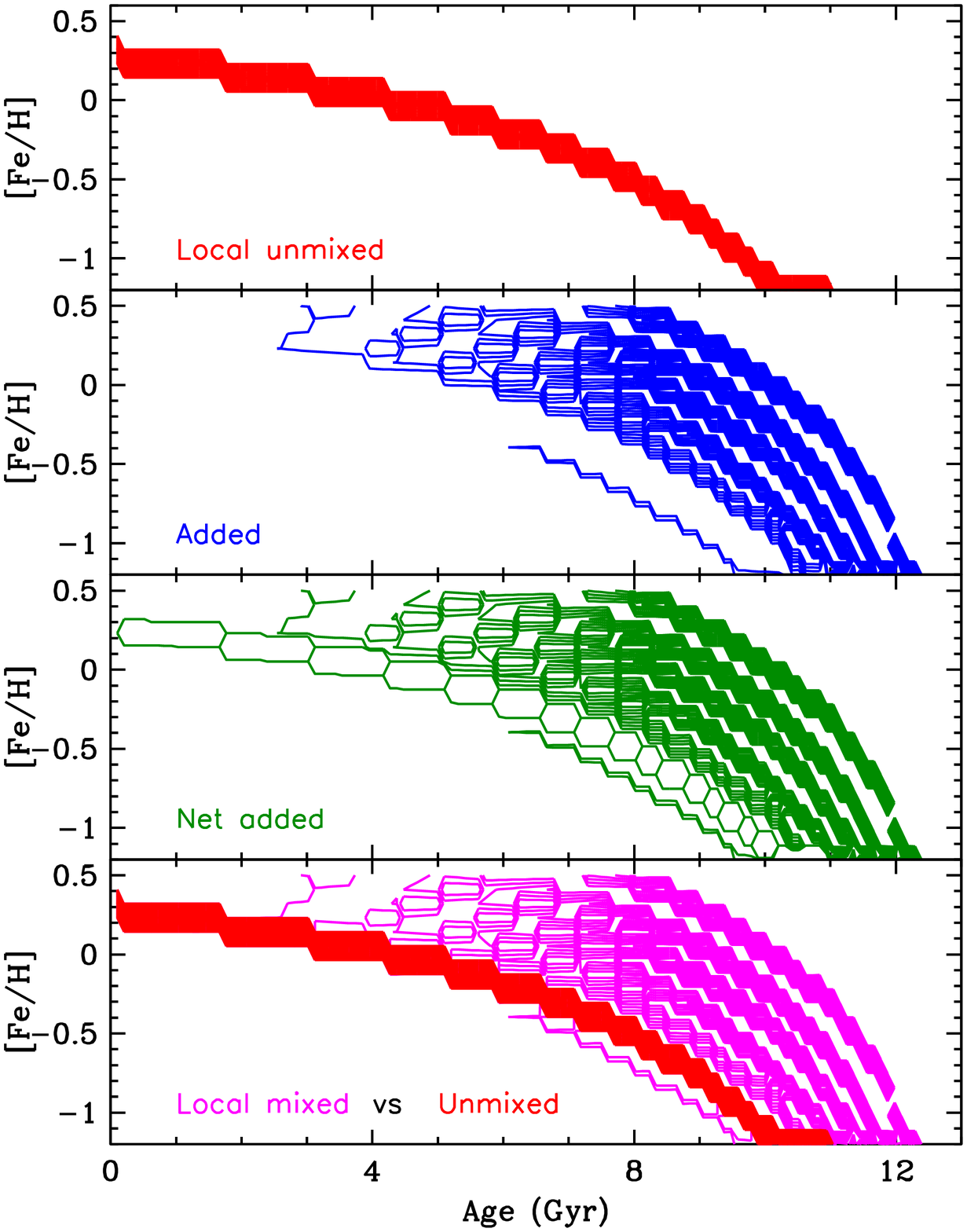}
\qquad
\includegraphics[width=0.46\textwidth]{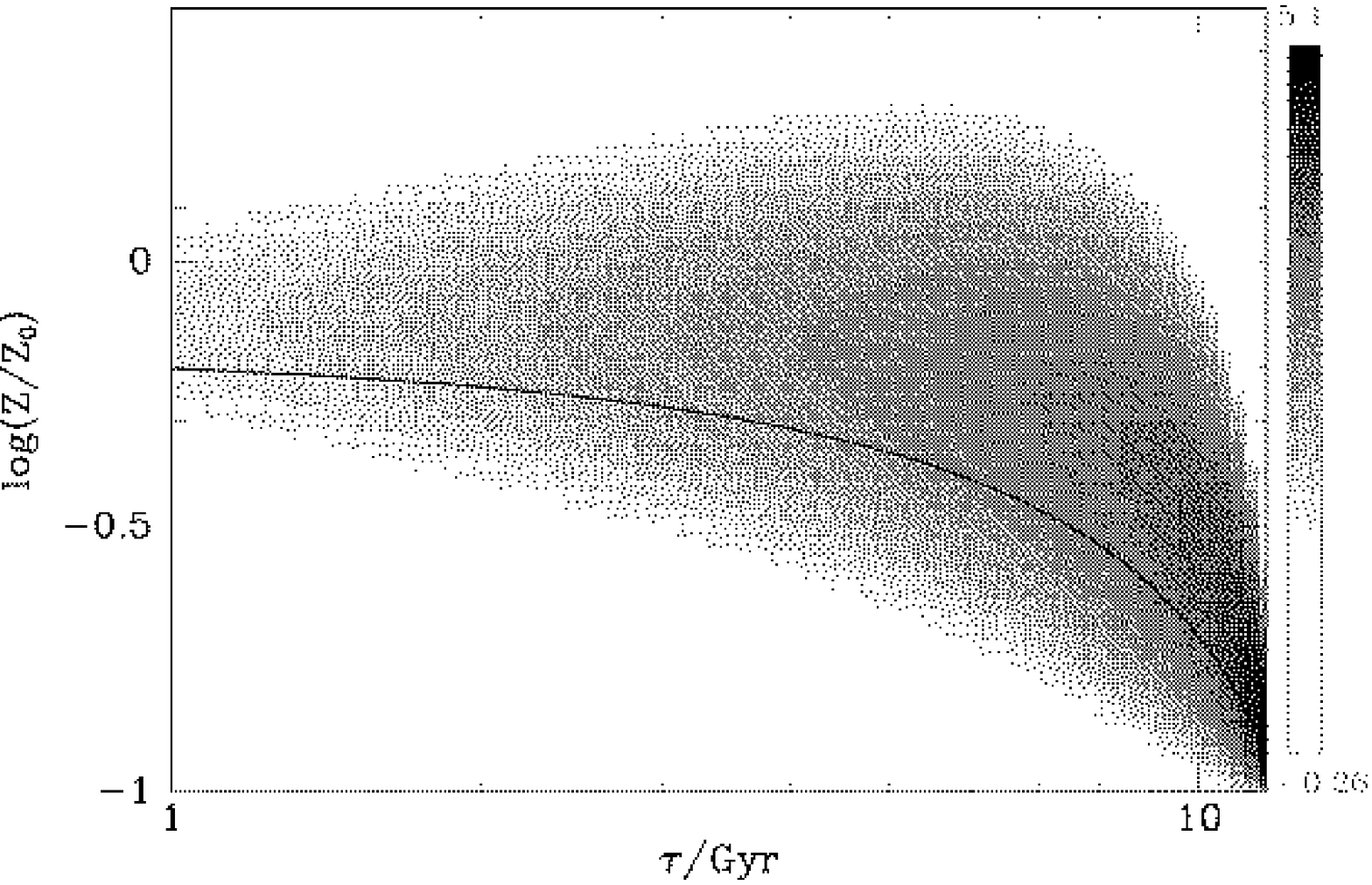}
\caption{{\bf Left:}  Impact of radial mixing on the local (R=8 kpc) age-metallicity relation accrding to the toy model
of the previous figure; from top to bottom: original age-metallicity relation, added material, net added material
(after accounting from removed matter) and total age-metallicity relation, with considerable scatter at old stellar ages. The
latter should be compared to the figure on the {\bf right}, from the toy model of Sellwood and Binney (2002).}
\end{figure}

The implications of radial mixing for the solar neighborhood appear in Fig. 6 (left panel), where
the age-metallicity relation appear for the locally born stars (top), for those brought from
other regions (middle) and for the total final sampl, i.e. original+added-removed (bottom).
A considerable scatter of the AMR is obtained, especially for old stars, confirming the finding
of SB02 with a simpler original model. It should be noticed, however, that in both models the
final result ressembles little to the stellar sample of the GKS, which displays i) a large
number of young stars with considerable metallicity scatter and ii) a flat early
AMR with a small upturn at late times.  The toy model explored here shows
a different trend, both concerning the scatter (important at early times) and the upturn (only at early times). This may imply two things: either i) the GKS sample is seriously biased, favouring 
excessively young stars, but also underestimating age errors for old stars, in which case it can hardly be used for comparison to  chemical evolution models (with our without mixing), or ii) the true AMR, both locally and galaxy-wide, is very different from the one resulting from standard models and adopted in the radial mixing models presented here.

\begin{figure}
\includegraphics[width=0.46\textwidth]{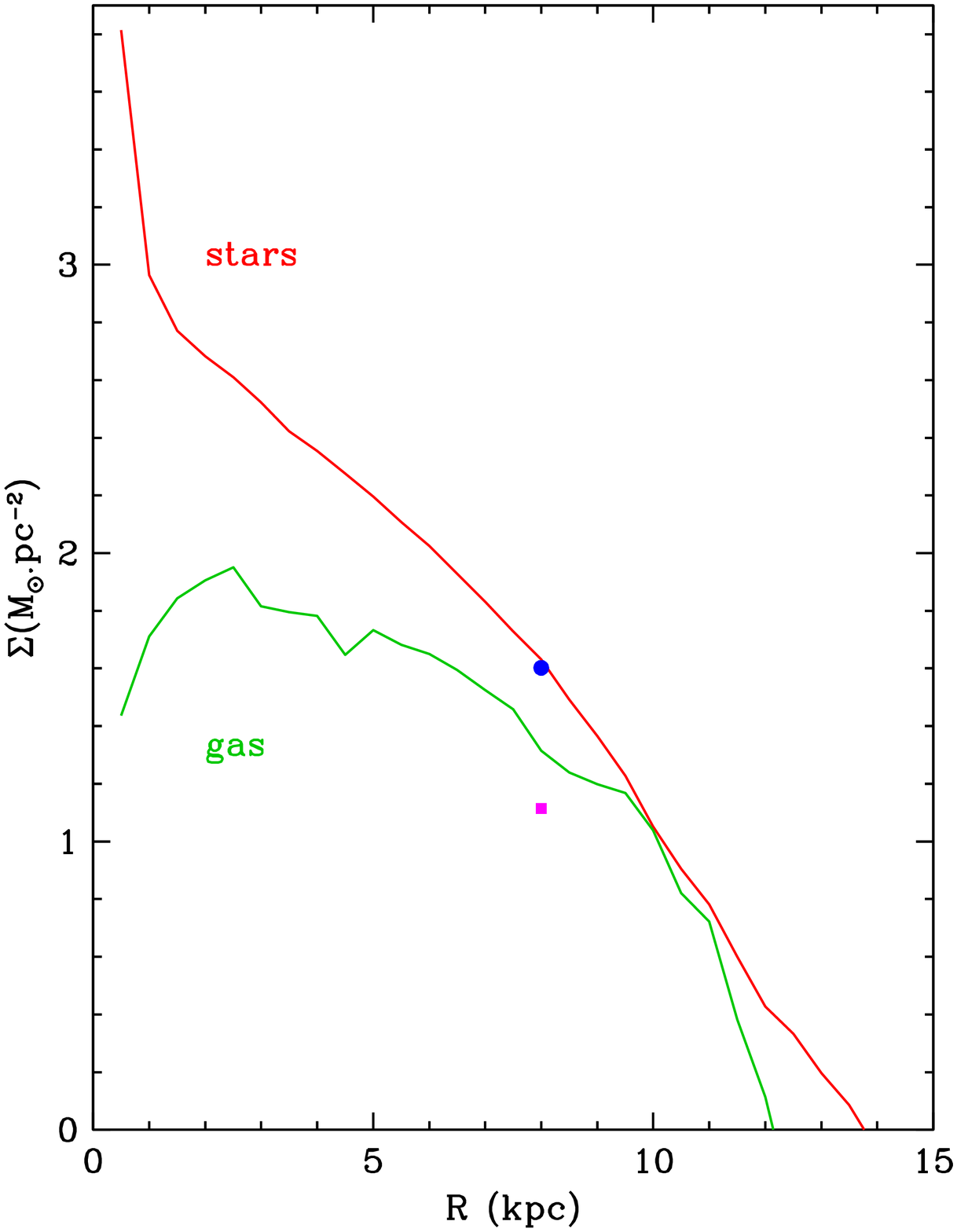}
\qquad
\includegraphics[width=0.46\textwidth]{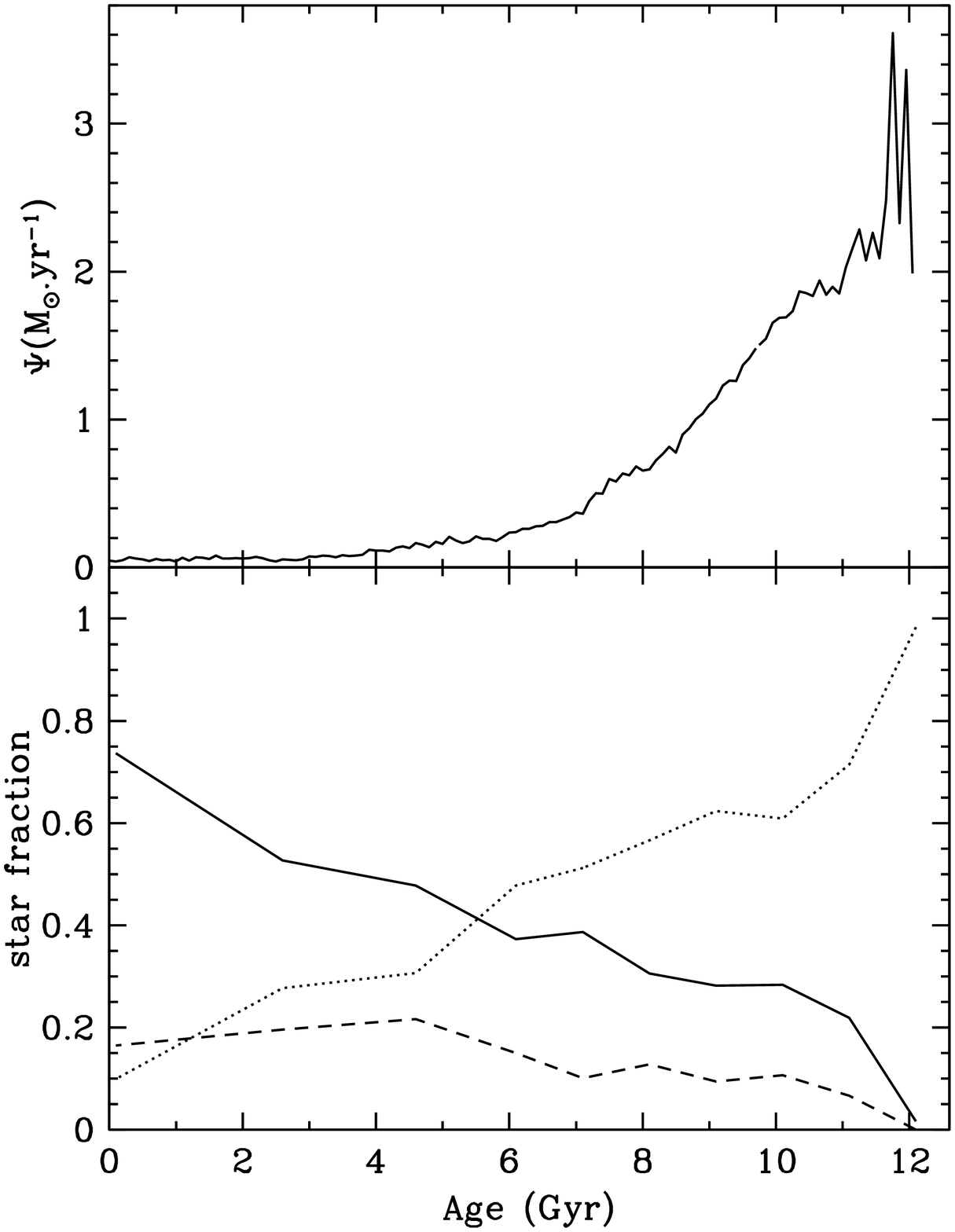}
\caption{{\bf Left:} Stellar and gaseous surface density profiles (log
 scale) of the N-body + SPH simulation of Heller et al. (2007). Points
 at 8 kpc represent corresponding quantities in the solar
 neighborhood. However, the simulation produces an early-type galaxy,
 as can be seen from the star formation rate vs. age (right top)
 and its results cannot be directly compared to the Milky way
 data. Nevertheless,considerable mixing of stars of all ages is found in
 all radial zones, as can be seen from the situation at 8 kpc (right
 bottom), where the fraction of locally born stars ({\it solid)}) is
 plotted against  age and compared to stars that migrated there from  the inner
 disk ({\it dotted}) and the outer disk ({\it dashed}).}
\end{figure}

A convincing case for radial mixing of stars in disks was recently presented in Roskar et al.
(2008), with high-resolution SPH + N-body simulations of an isolated  galaxy 
growing from dissipational collapse of gas embedded in a spherical dark matter halo.
They find a break in star formation in the outer disk (due to the unaivalability of
 sufficiently cooled  gas in those regions) which moves gradually outwards. Despite 
 that break, they find a large number of stars outside the final break radius, 
 which are scattered there from the inner disk on  nearly circular orbits by collisions with
 spiral arms.
 
 We explored the  issue of radial mixing by analysing the results of another 
 SPH + N-body simulation, performed  by Heller et a. (2007). It concerns a 
 galaxy growing in an isolated  dark matter halo and developing final stellar 
 and gaseous profils not unlike the corrswponding ones of the Milky Way
 (Fig. 7 left panel). However, it is an early type galaxy, since most of
  its star formation occurs early on (top right panel in Fig. 7), and the results 
  cannot be compared in detail to observations of the Milky Way. Considerable mixing of
  stars is found at $R$=8 kpc, mostly originating from the inner disk (Fig. 7 right bottom); in fact, most of the local stars of age $>$7 Gyr come from the inner disk
  in that similation, while the fraction of stars coming in from the outer disk never exceeds 20\%. The complete analysis of these results and their impact on the 
  metallicity distribution are presented in Bellil et al. (2008, in preparation). We expect the effect
  of radial mixing on local properties to be smaller for a MW-type galaxy, because of its slower
  star formation, which leaves less time for stellar transport. 

A  last, but important point, concerning radial mixing: independently of theoretically appealing arguments, observations will ultimately determine its importance. In that respect, a strong 
constraint to consider is the resulting [O/Fe] ratio as a function of metallicity: in the
solar neighborhood, that ratio displays very little scatter (if any at all). If a large number
of local stars originated in inner disk regions (with large early metallicities {\it and} large O/Fe 
ratios, because of the early absence of Fe producing SNIa), one would expect to find a substantial scatter  in that ratio in the local sample.
We are currently working to quantify that effect and
constrain the extent of radial mixing in the MW.

{}


\begin{thebibliography}{}

\bibitem[] { } Bekki, K., Chiba, M., 2001, ApJ 558, 666

\bibitem[] { } Bell, E., Zuker, D, Belokurov, V., 2008, ApJ 680, 295

\bibitem[] { } Boissier, S., Prantzos, N., 1999, MNRAS 307, 857

\bibitem[]{} Edvardsson, B., Andersen, J., Gustaffson B., et al., 1993, A\&A 275, 101

\bibitem[]{} Dekel, A., Woo, J.,  2003, MNRAS 344, 1131 

\bibitem[]{} Diemand, J., Kuhlen, M., Madau, P., 2007, ApJ 667, 859 

\bibitem[]{} Feltzing, S., Holmberg, J., Hurley, J. R., A\&A 377, 911

\bibitem[]{} Font, A., Johnston, K., Bullock, J., Robertson, B., 2006, ApJ, 638, 585

\bibitem[]{} Giocoli, C., Pieri, L.,  Tormen, G. 2008, MNRAS 387, 689

\bibitem[] { } Goswami, A., Prantzos, N., 2000, A\&A 359, 151

\bibitem[] { } Hartwick, F., 1976, ApJ 209, 418

\bibitem[]{} Haywood, M., 2006, MNRAS 371, 176

\bibitem[]{} Haywood, M., 2008, MNRAS 388, 1175

\bibitem[]{} Heller, C., Shlosman, I., Athanassoula, E., 2007, ApJ 671, 226

\bibitem[]{} Helmi, A., Irwin, M., Tolstoy, E., et al., 2006, ApJ 651, L121

\bibitem[]{} Holmberg, J., Norstr\" om, B., Andersen, J., 2007, A\&A 475, 519

\bibitem[]{} Kroupa, P., 2002, Science 295, 82

\bibitem[]{} Lepine, J. R.. D., Acharova, I. A., Mishurov, Y. N., 2003, ApJ 589, 210

\bibitem[]{} Mayor, M., 1974, A\&A 32, 321

\bibitem[]{} Nordstr\" om, B., Mayor, M., Andersen, J., et al., AA418, 989 

 \bibitem[] { } Pagel B., 1997, ``Nucleosynthesis and galactic chemical 
               evolution''              (Cambidge University Press)

\bibitem[]{} Prantzos, N. 2003, A\&A 404, 211

\bibitem[]{} Prantzos, N. 2007a, in "Stellar Nucleosynthesis: 50 years
	  after B2FH", C. Charbonnel and J.P. Zahn (Eds.), EAS
	  publications Series  (arXiv:0709.0833)

\bibitem[]{} Prantzos, N. 2007b, in "CRAL-2006: Chemodynamics, from first stars
	  to local galaxies", E. Emsellem et al. (Eds.), EAS
	  publications Series  Vol. 24, p. 3 (arXiv:astro-ph/0611476)

\bibitem[]{} Prantzos, N. 2008, A\&A in press (arXiv:0807.1502)

\bibitem[]{} Roskar,R., Debattista, V., Stinson, G., et al., 2008, ApJ 675, L65

\bibitem[] { } Ryan S., Norris J., 1991,  AJ 101, 1865

\bibitem[]{} Salvadori, S, Schneider, R., Ferrara, A., 2007, MNRAS 381, 647

\bibitem[]{} Salvadori, S, Ferrara, A., Schneider, R., 2008, MNRAS 361, 348

\bibitem[]{} Scanapieco, E., Broadhurst, T., 2001, ApJ 550, L39

\bibitem[]{} Sellwood, J., Binney, J.,  2002, MNRAS 336, 785 

\bibitem[]{} Shetrone, M., Cot\'e, P., Sargent, W., 2001, ApJ 548, 592

\bibitem[]{} Soubiran, C., Girard, P., 2005, A\&A 438, 139

\bibitem[]{} Soubiran, C..,  Bienaym\'e, O., Mishenina, T. V., Kovtyukh, V. V. 2008, A\&A 480, 91

\bibitem[]{} Tumlinson, J., 2006, ApJ  641, 1


\bibitem[]{} Twarog, B. A., 1980, ApJ 242, 242

\bibitem[]{} Venn, K., Irwin, M., Shetrone, M., et al., 2004, ApJ 128, 1177

\bibitem[]{} Wielen, R., Fuchs, B., Dettbarn, C., 1996, A\&A 314, 438


\end{thebibliography}
\end{document}